\documentclass[]{spie}  %>>> use for US letter paper
\usepackage[]{graphicx}

\title{Reduction Algorithms for the Multiband Imaging Photometer for
   Spitzer: 6 Months of Flight Data}  

\author{
K. D. Gordon, C. W. Engelbracht, J. Muzerolle,
J. A. Stansberry, K. A. Misselt, J.~E.~Morrison, G. Rieke, J. Cadien,
E. T. Young, H. Dole, D. M. Kelly, A. Alonso-Herrero, E. Egami,
K.~Y.~L.~Su, C. Papovich, P. S. Smith, D. C. Hines, M. J. Rieke,
M. Blaylock, P.~G.~P\'erez-Gonz\'alez, E. Le Floc'h, J. Hinz
\skiplinehalf
Steward Observatory, University of Arizona, Tucson, AZ, 85721, USA
\skiplinehalf
and
\skiplinehalf
W. B. Latter, T. Hesselroth, D. T. Frayer, A. Noriega-Crespo, F. J. Masci,
D. L. Padgett
\skiplinehalf
Spitzer Science Center, Pasadena, CA, 91125, USA
}

\authorinfo{Send correspondence to K.\
   D.\ Gordon, Email: kgordon@as.arizona.edu}
\begin{document} 
\maketitle 

%%%%%%%%%%%%%%%%%%%%%%%%%%%%%%%%%%%%%%%%%%%%%%%%%%%%%%%%%%%%% 
\begin{abstract}
The first six months of flight data from the Multiband Imaging
Photometer for Spitzer (MIPS) were used to test MIPS reduction
algorithms based on extensive preflight laboratory data and modeling.
The underlying approach for the preflight algorithms has been found to
be sound, but some modifications have improved the performance.  The
main changes are scan mirror dependent flat fields at 24~$\mu$m, hand
processing to remove the time dependent stim flash latents and
fast/slow response variations at 70~$\mu$m, and the use of asteroids
and other sources instead of stars for flux calibration at 160~$\mu$m
due to a blue ``leak.''  The photometric accuracy of flux measurpements
is currently 5\%, 10\%, and 20\% at 24, 70, and 160~$\mu$m,
respectively.  These numbers are expected to improve as more flight
data are analyzed and data reduction algorithms refined.
\end{abstract}

%>>>> Include a list of keywords after the abstract 

%\keywords{Need some.}

%%%%%%%%%%%%%%%%%%%%%%%%%%%%%%%%%%%%%%%%%%%%%%%%%%%%%%%%%%%%%
\section{INTRODUCTION}
\label{sect:intro}  % \label{} allows reference to this section

The Multiband Imaging Photometer for Spitzer (MIPS) provides the
far-infrared capabilities for the Spitzer Space Telescope.  MIPS
provides images at 24, 70, and 160~$\mu$m with well sampled
point-spread-functions.  In addition, low resolution spectra from 52
to 100~$\mu$m can be obtained using the Spectral Energy Distribution
mode.  The detectors at 24, 70, and 160~$\mu$m are a 128x128 Si:As BIB
array, a 32x32 Ge:Ga array, and a 20x2 stressed Ge:Ga array.  The MIPS
instrument is described by Rieke et al.\cite{Rieke04a} and Rieke et
al.\cite{Rieke04b} and the Spitzer Space Telescope by Werner et
al.\cite{Werner04}.  The reduction of MIPS data is the subject of this
paper, specifically the modifications needed to preflight
algorithms\cite{Gordon04} after analysis of the first six months of
flight data.  As the analysis of flight data is ongoing, this paper is
necessarily a progress report.

%%%%%%%%%%%%%%%%%%%%%%%%%%%%%%%%%%%%%%%%%%%%%%%%%%%%%%%%%%%%%
\section{SUMMARY OF PREFLIGHT REDUCTION ALGORITHMS} 

Extensive efforts before the launch of Spitzer were made to prepare
reduction algorithms for all three MIPS arrays. The results of this
work are presented by Gordon et al.\cite{Gordon04} and summarized in
this section.

\begin{figure}
\begin{center}
\begin{tabular}{c}
\includegraphics[height=17cm]{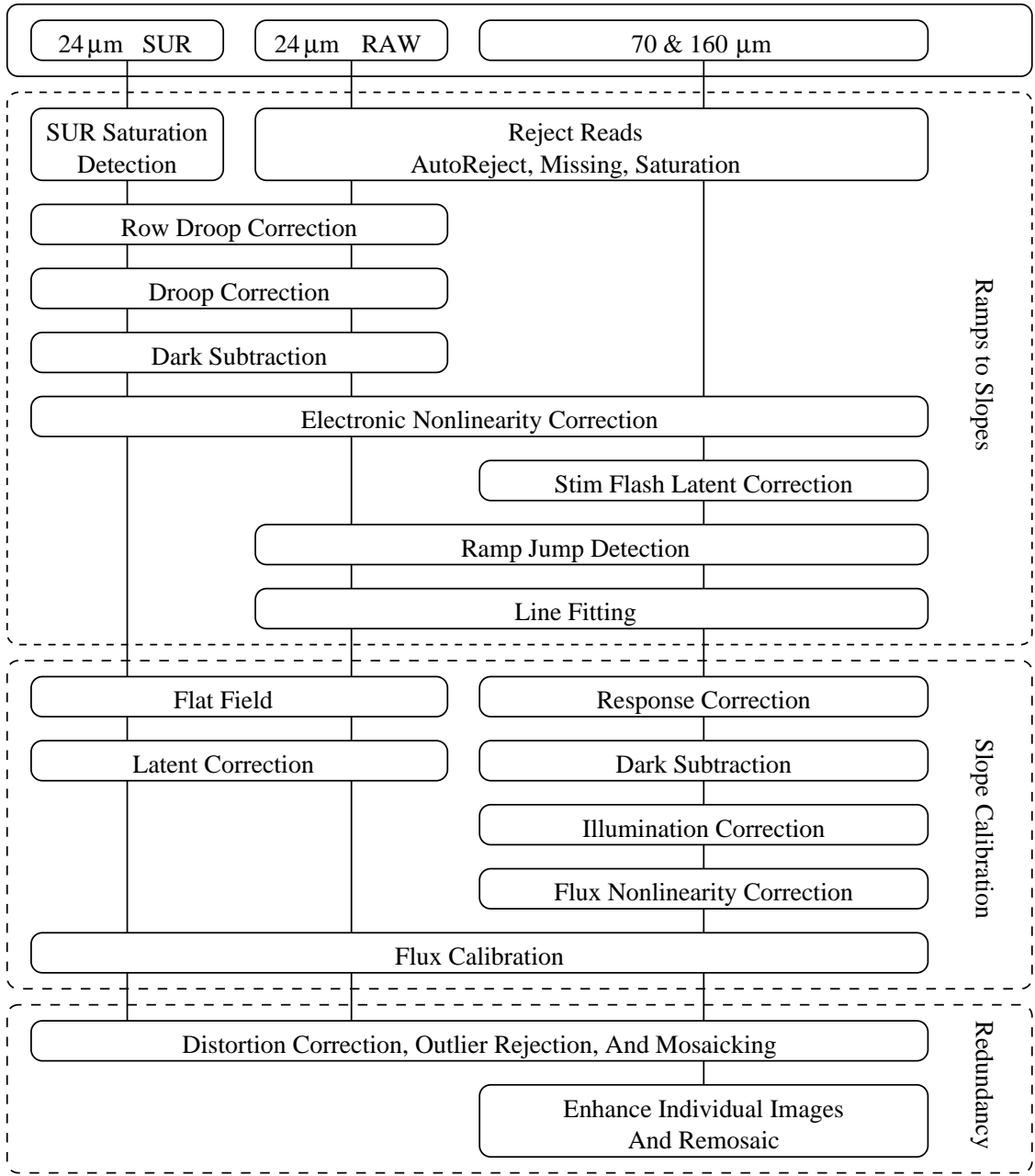}
\end{tabular}
\end{center}
\caption[dat_flowchart] 
%>>>> use \label inside caption to get Fig. number with \ref{}
{ \label{fig:dat_flowchart} Graphical representations of the flow of
reduction of MIPS data based on preflight expectations.}
\end{figure} 

Figure~\ref{fig:dat_flowchart} gives a pictorial representation of the
preflight algorithms split into three stages.  The first stage
(``Ramps to Slopes'') transforms the non-destructive samples in an
integration ramp into a slope measurement while performing corrections
with time scales on order of the exposure time.  The second stage
(``Slope Calibration'') calibrates the raw slopes to remove
instrumental effects with time scales longer than an individual
exposure time.  Finally, the third stage (``Redundancy'') takes a set
of exposures and combines them into a single mosaicked image while
correcting for optical distortions and rejecting outliers.  This third
stage was the hardest to define fully with only preflight data,
especially how to make full use of the redundancy in the data-taking
to improve the removal of instrument signatures and data artifacts.

All three arrays are non-destructively read out continuously at
intervals of either 1/2 or 1/8 of a second for the 24 and
70/160~$\mu$m arrays, respectively.  After the exposure is complete,
the arrays are reset and the charge starts building up again.  The
majority of 24~$\mu$m flight data is taken in sample-up-the-ramp (SUR)
mode, to limit the total amount of data that must be stored on-board
the spacecraft. In this mode, a slope is fit onboard Spitzer and the
resulting slope measurement downlinked.  In addition, the difference
between the first two reads (1st difference) is also downlinked to
increase the dynamic range.  Some 24~$\mu$m engineering data and all
70 and 160~$\mu$m data are taken in RAW mode, under which all the
non-destructive reads are downlinked. In these cases, the basic unit
of data is a well-sampled charge accumulation ramp, and slopes are
fitted on the ground after preliminary processing steps such as
rejecting invalid data and making linearity corrections.

While the three MIPS arrays share some reduction steps, there is a
natural division between the reduction of the Si:As 24~$\mu$m array
data and that from the Ge:Ga 70 and 160~$\mu$m arrays.

For 24~$\mu$m SUR mode data, the first step in the ``Ramps to Slopes''
stage is the detection of saturation, which is done by comparing the
slope and 1st difference values.  Next, row droop and droop
corrections are applied.  Droop is an electronic effect that raises
the overall output level of the whole array by an amount proportional
to the total charge on the array.  Row droop is similar but only
applies along rows.  The very small dark signal is subtracted and
nonlinearities (up to 14\%) due to the electronics corrected based
only on the overall slope.  When 24~$\mu$m RAW mode data are reduced,
similar steps are applied but on the full ramp data instead of the
slopes.  In addition, having the ramp means that saturation detection
is cleaner and cosmic rays (ramp jumps) can be detected and
removed. The final step is then fitting the ramp to produce the slope
measurement.  The remaining steps are identical for both SUR and RAW
mode data. In the ``Slope Calibration'' stage, the flat field is
applied, source latent images are corrected and the flux calibration
applied.  Finally, the ``Redundancy'' stage creates the final mosaics
after correcting for distortions and rejecting outliers.

For the 70 and 160~$\mu$m data, the first step in the ``Ramps to
Slopes'' stage is to reject bad reads in the charge ramps.  These bad
reads can be due to reset pulse transients (autoreject), missing data,
or saturation.  Next, the small (1-2\%) electronic nonlinearities are
corrected and the stim flash latents removed.  The stimulators are
used to track the changing response of the Ge:Ga detectors and are
flashed about every two minutes.  These bright flashes leave latent
images, but with amplitudes less than a percent or two and time
constants less than 20 seconds.  Next, jumps in the ramps due to
cosmic rays or electronics (readout jumps) are detected.  Finally, the
ramp segments are fit to the valid segments of data and the slope
measurement determined.  The ``Slope Calibration'' stage starts by
correcting the changing response of the detectors by dividing by the
stim flashes.  The dark, also calibrated by dividing by the stim
flashes, is subtracted.  The illumination pattern of the
telescope/instrument and the stim flash is removed using an
illumination correction determined using flat sky observations (like a
traditional flat field).  The bulk Ge:Ga photoconductors are subject
to flux dependent signal nonlinearities. These effects are corrected
on a pixel-by-pixel basis using extensive calibration observations.
Finally, the flux calibration based on stellar observations is
applied.  In the ``Redundancy'' stage, optical distortions are removed
and outliers rejected using the redundancy inherent in the basic unit
of observations.  Finally, the redundancy is used to search for
residual instrumental signatures to remove them and then the data are
remosaicked.  The approaches for this final step are only now being
defined, as flight data are analyzed. We anticipate that improvements
will be made throughout the mission.

%%%%%%%%%%%%%%%%%%%%%%%%%%%%%%%%%%%%%%%%%%%%%%%%%%%%%%%%%%%%%
\section{FLIGHT DATA} 

The preflight algorithms reflect extensive analysis of laboratory data
on the flight or similar arrays and theoretical modeling of Ge:Ga
detectors.  Nevertheless, it was always expected that the realities of
data taken in flight would require modification of these algorithms.
This paper details the changes resulting from the continuing analysis
of MIPS flight data during the three months of instrument
commissioning data (In Orbit Checkout/Science Verification) and
roughly three months of nominal science operations.

\begin{figure}
\begin{center}
\begin{tabular}{c}
\includegraphics[height=18cm]{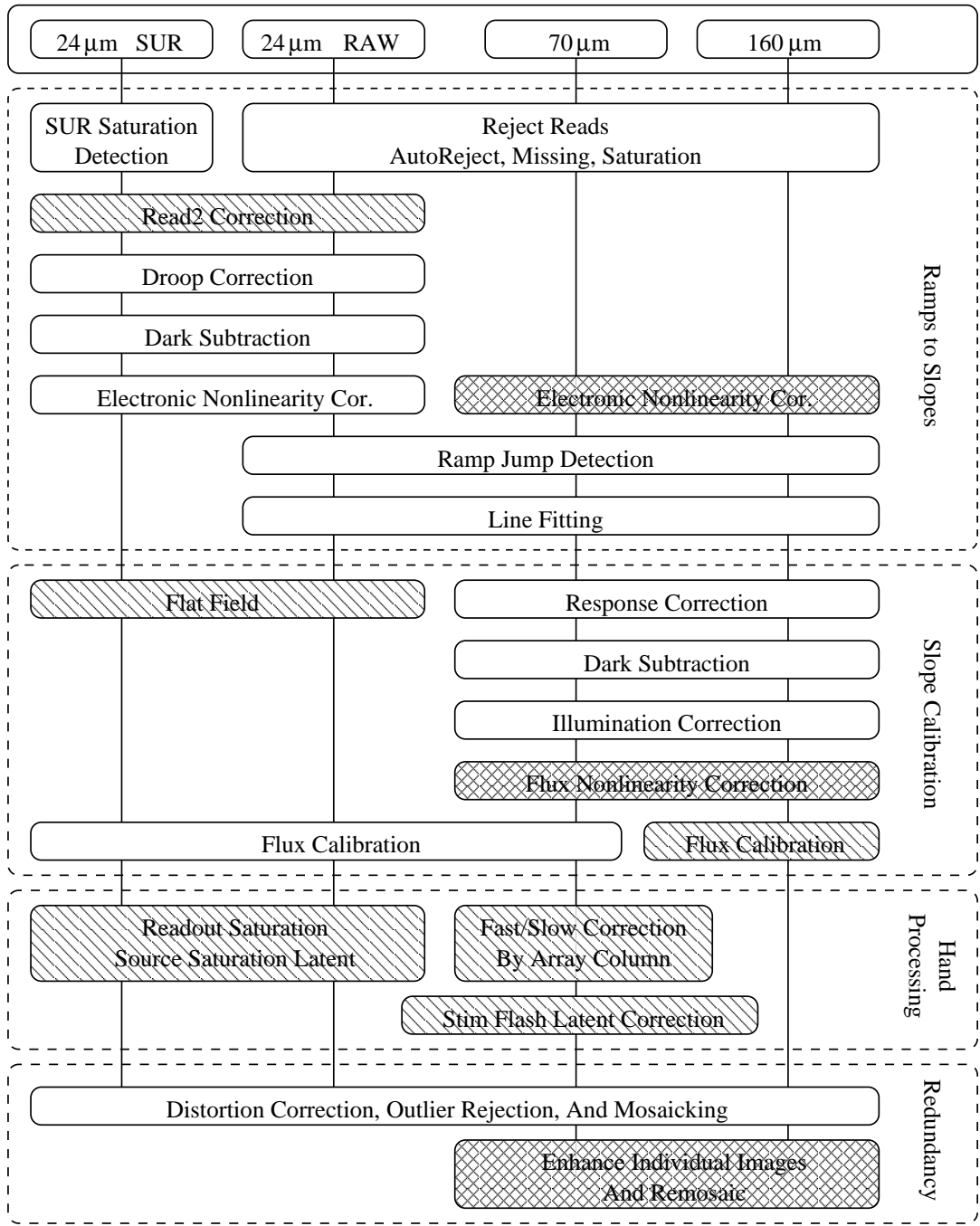}
\end{tabular}
\end{center}
\caption[dat_flowchart_flight] 
%>>>> use \label inside caption to get Fig. number with \ref{}
{ \label{fig:dat_flowchart_flight} Graphical representations of 
the flow of reduction of MIPS data based on preliminary analysis of 
flight data.  Left leaning hatching denotes steps which are
significantly changed from preflight expectations and left/right
hatching means steps which have not been implemented for flight data
yet.} 
\end{figure} 

The current state of knowledge of the best MIPS data reduction
algorithms is given in pictorial form in
Figure.~\ref{fig:dat_flowchart_flight}.  The details of the
modifications from Figure~\ref{fig:dat_flowchart} are discussed in the
following sections for each array in turn.  The largest change in the
reduction algorithms is the addition of a ``Hand Processing'' stage
between the ``Slope Calibration'' and ``Redundancy'' stages.  The
``Hand Processing'' may disappear in the future after sufficient
knowledge and testing occurs to provide automatic versions of these
steps.

\subsection{24 $\mu$m}

\begin{figure}
\begin{center}
\begin{tabular}{c}
\includegraphics[height=7cm]{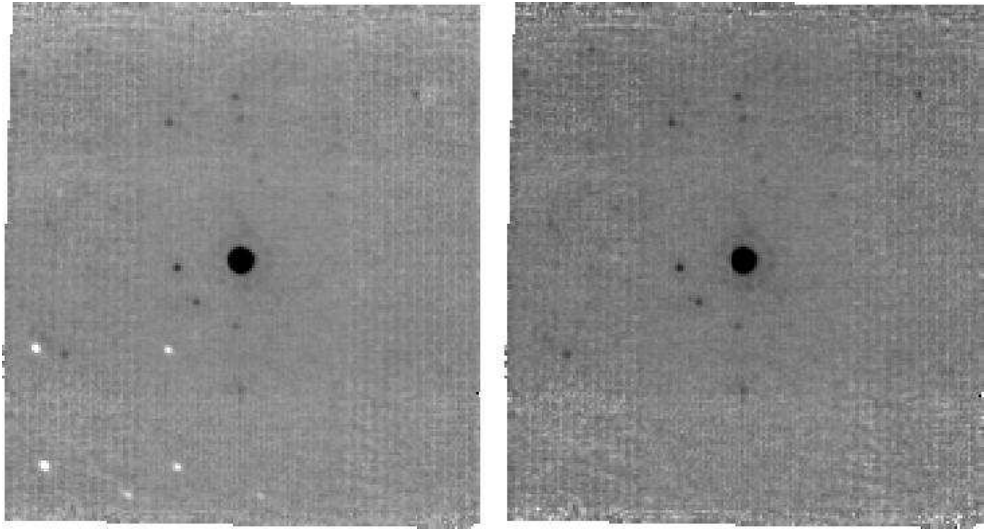}
\end{tabular}
\end{center}
\caption[smd_flat_24] 
{ \label{fig:smd_flat_24} The importance of scan mirror dependent
flats for reduction of 24~$\mu$m data is shown.  The left panel has a
single scan mirror independent flat applied and the right panel has
scan mirror dependent flats applied.  Both panels are the result of
mosaicking a set of images taken in photometry mode.  Notice the four
white points in a grid pattern in the left panel which are not present
in the right panel.  Other similar, weaker features can also be seen
in the left, but not right panel.} 
\end{figure}

As expected, the 24~$\mu$m reduction algorithms required only minor
modifications to produce excellent, well calibrated 24~$\mu$m images.
The most significant change required creating flat fields for each
scan mirror position used instead of a single flat field.  Small
particles were deposited on the pick-off mirror, probably during
launch.  The light they block leads to spots in the images whose
positions are dependent on the scan mirror angle.  By reducing the
24~$\mu$m data with scan mirror dependent flat fields, the spots are
cleanly removed as can be seen in Figure~\ref{fig:smd_flat_24}.

\begin{figure}
\begin{center}
\begin{tabular}{c}
\includegraphics[height=5.6cm]{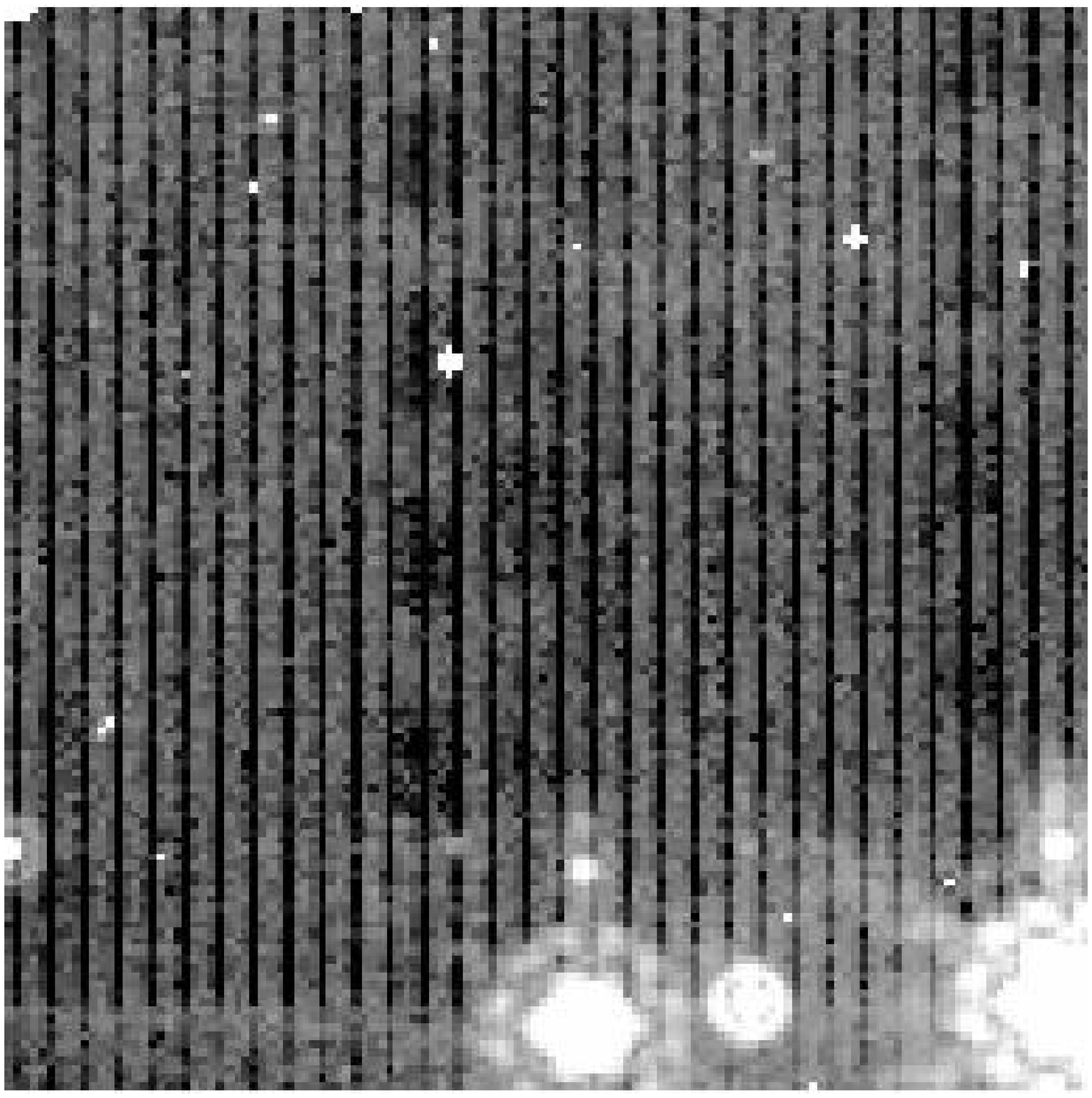}
\includegraphics[height=5.6cm]{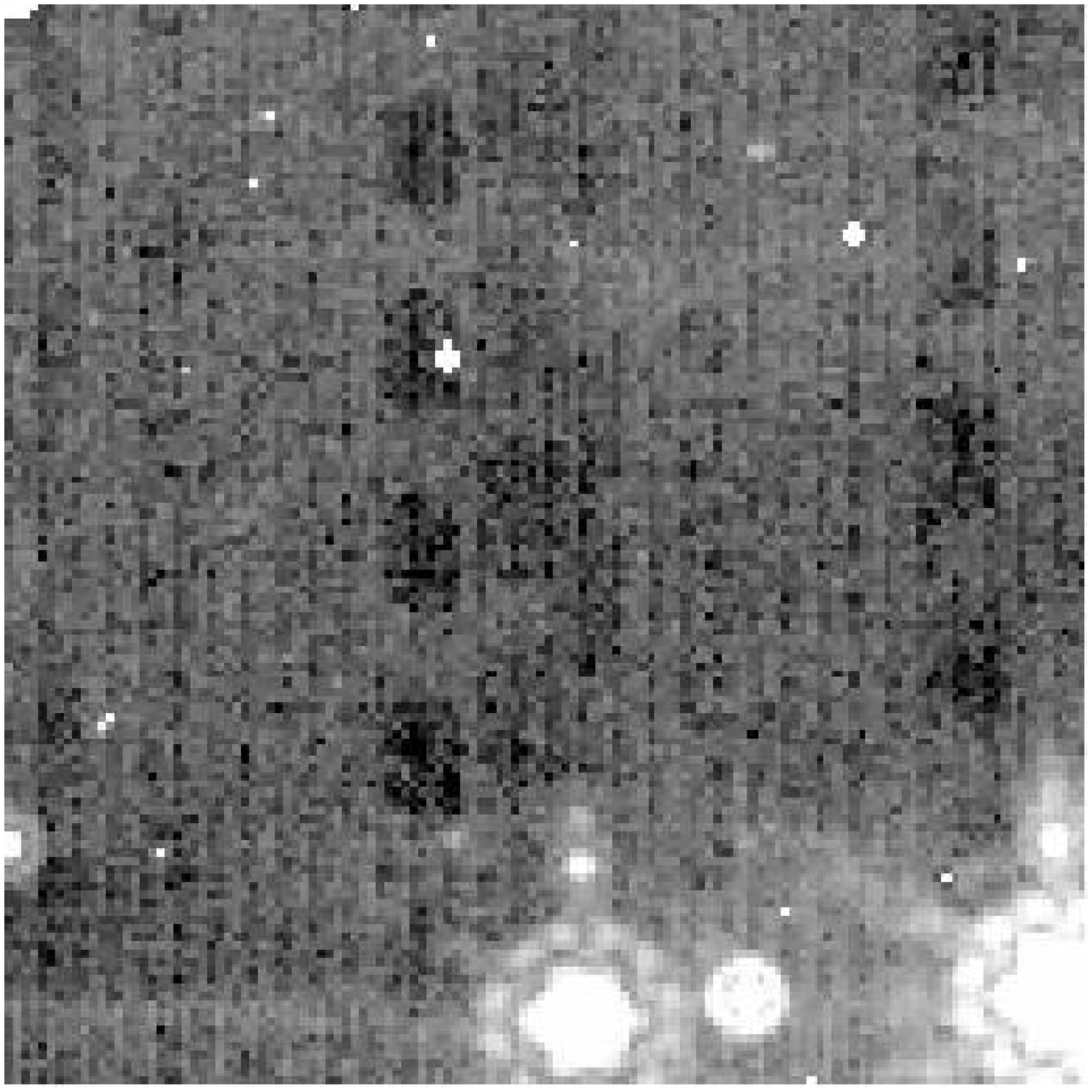}
\includegraphics[height=5.6cm]{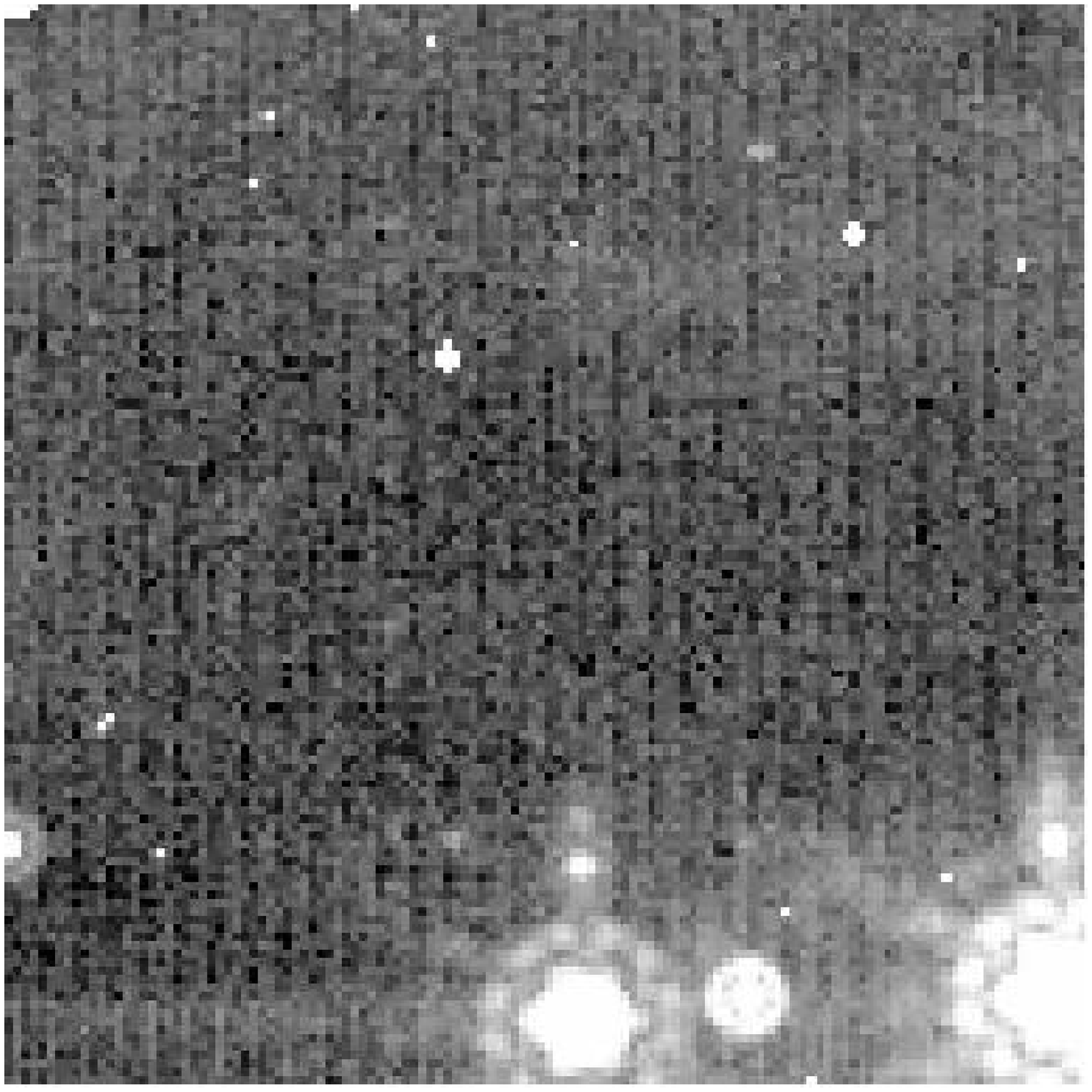}
\end{tabular}
\end{center}
\caption[sat_fig_24] 
{ \label{fig:sat_fig_24} The readout saturation and saturating source
latent are shown in the left panel.  The center panel has the readout
saturation correction applied.  The right panel has the readout and
saturation source saturations removed. }
\end{figure} 

The Row Droop effect was not seen in the flight data and the
correction for this effect has been removed.  In the place of this
step, a new correction for what is termed the ``Read2'' effect
has been inserted.  The ``Read2'' effect describes the bias introduced
into the SUR mode slope measurement by a small additive offset to the
second read of each 24~$\mu$m ramp (the first read is automatically
ignored due to known reset transients).  This additive offset varies
across the array and has been empirically calibrated using flight RAW
mode data.

The Latent Correction has been removed from the ``Slope Calibration''
stage as the characterization of this correction has proven to be more
difficult than anticipated.  While such latents are not a major
concern, it may be possible to reinstate this correction at a later
date once enough data are accumulated to provide a good calibration.

There are two effects associated with saturation that are in the
``Hand Processing'' category.  The first is caused by a hard
saturating cosmic ray or astronomical source that suppresses the
output of a single readout; it is termed readout saturation.  As there
are only four readouts for the 24~$\mu$m array, this has the
appearance of depressing every fourth column.  The readout saturation
can be corrected for most affected images by using the other three
readouts to provide the correct overall level.  The second is seen
after a saturating object and results in a source latent appearing for
many images afterwards.  The long life of this latent means that it is
possible to create a correction for the affected images on a
case-by-case basis, but not for images which show significant complex
structure.

\begin{figure}
\begin{center}
\begin{tabular}{c}
\includegraphics[height=7cm]{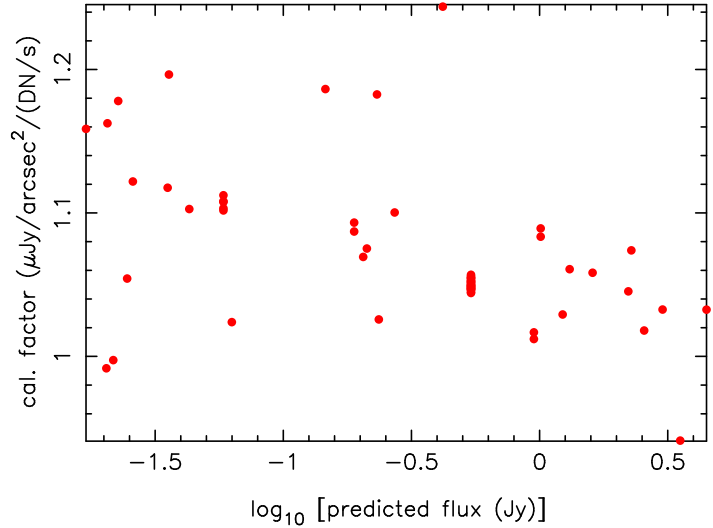}
\end{tabular}
\end{center}
\caption[cal_factor_24] 
{ \label{fig:cal_factor_24} The 24~$\mu$m calibration factor measured
for the calibration star measurements (minus two high and two low
outliers) is plotted. } 
\end{figure} 

The overall behavior of the 24~$\mu$m array only can be described as
excellent.  Mosaics made from many individual images are clean with
few, if any, instrumental residuals present.  Photometry repeats to
1\% or better and the absolute calibration is better than 5\%
determined using observations of solar analog, A, and K giant stars
(see Figure~\ref{fig:cal_factor_24}).

\subsection{70 $\mu$m}

The reduction of the 70~$\mu$m data has benefited more than the other
two arrays from the ``Hand Processing'' stage.  This mainly has been
the result of the discovery of significant time evolution of the stim
flash latents and fast/slow response variations seen in flight data.

\begin{figure}
\begin{center}
\begin{tabular}{c}
\includegraphics[height=5.5cm]{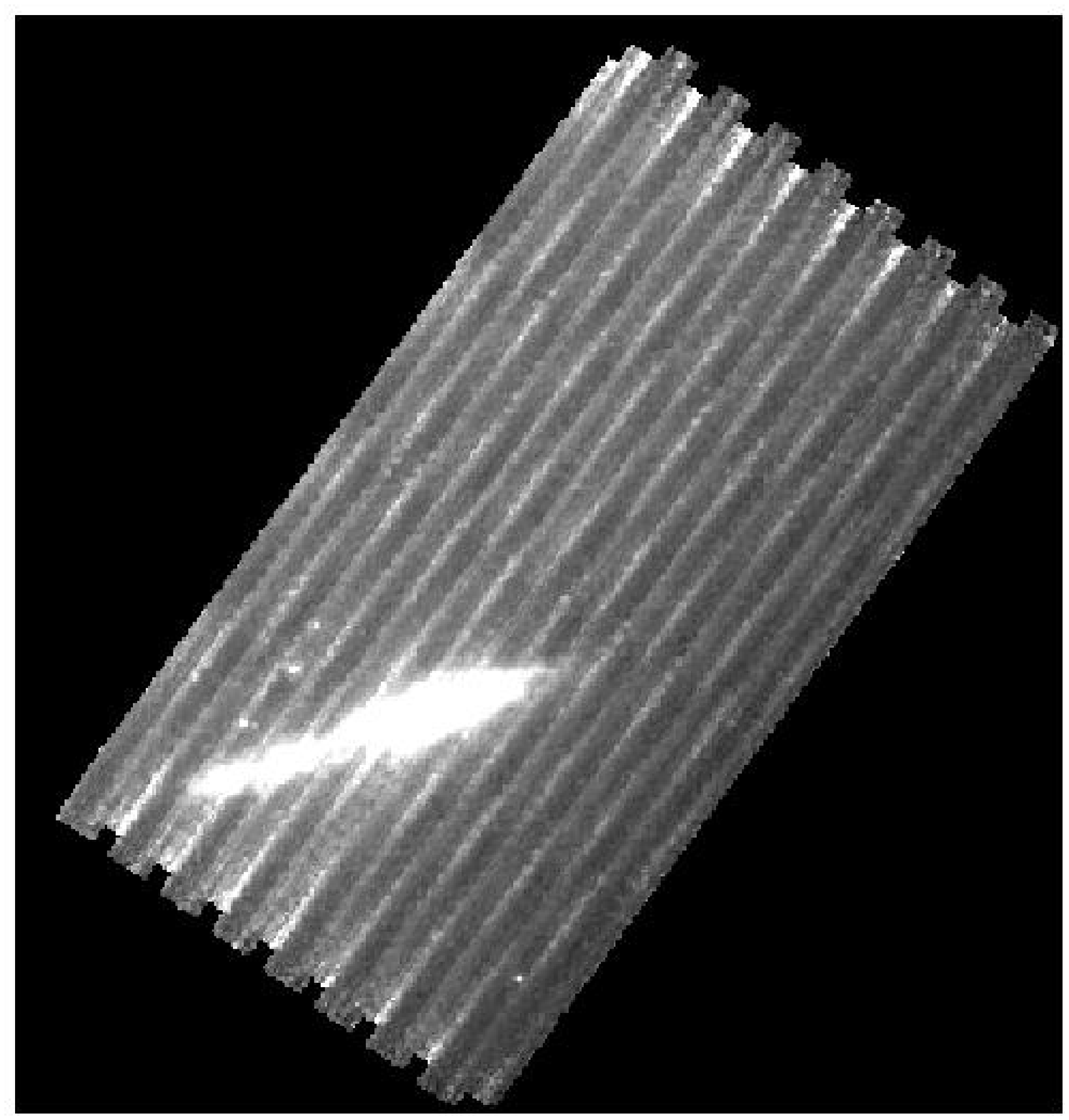}
\includegraphics[height=5.5cm]{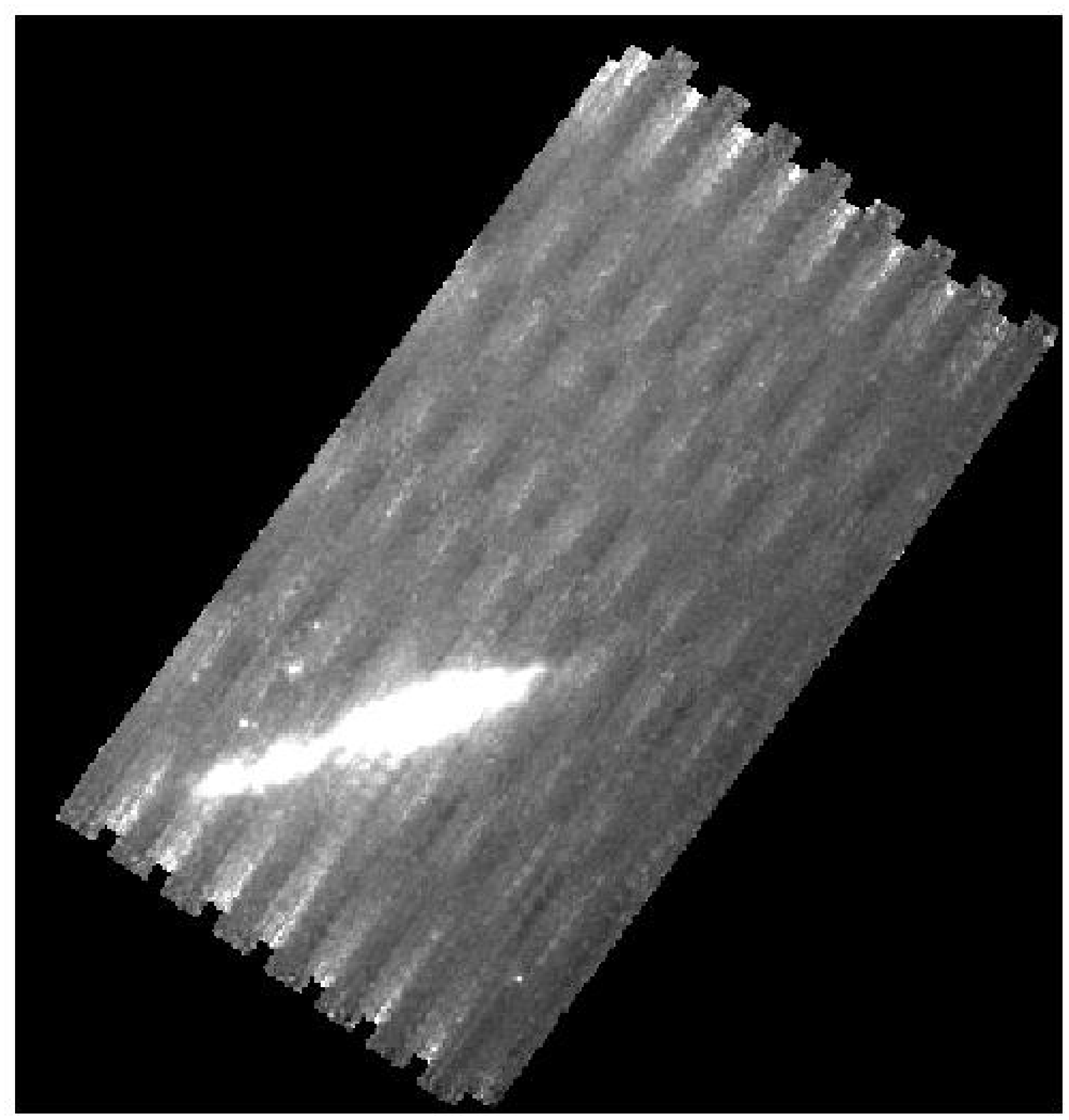}
\includegraphics[height=5.5cm]{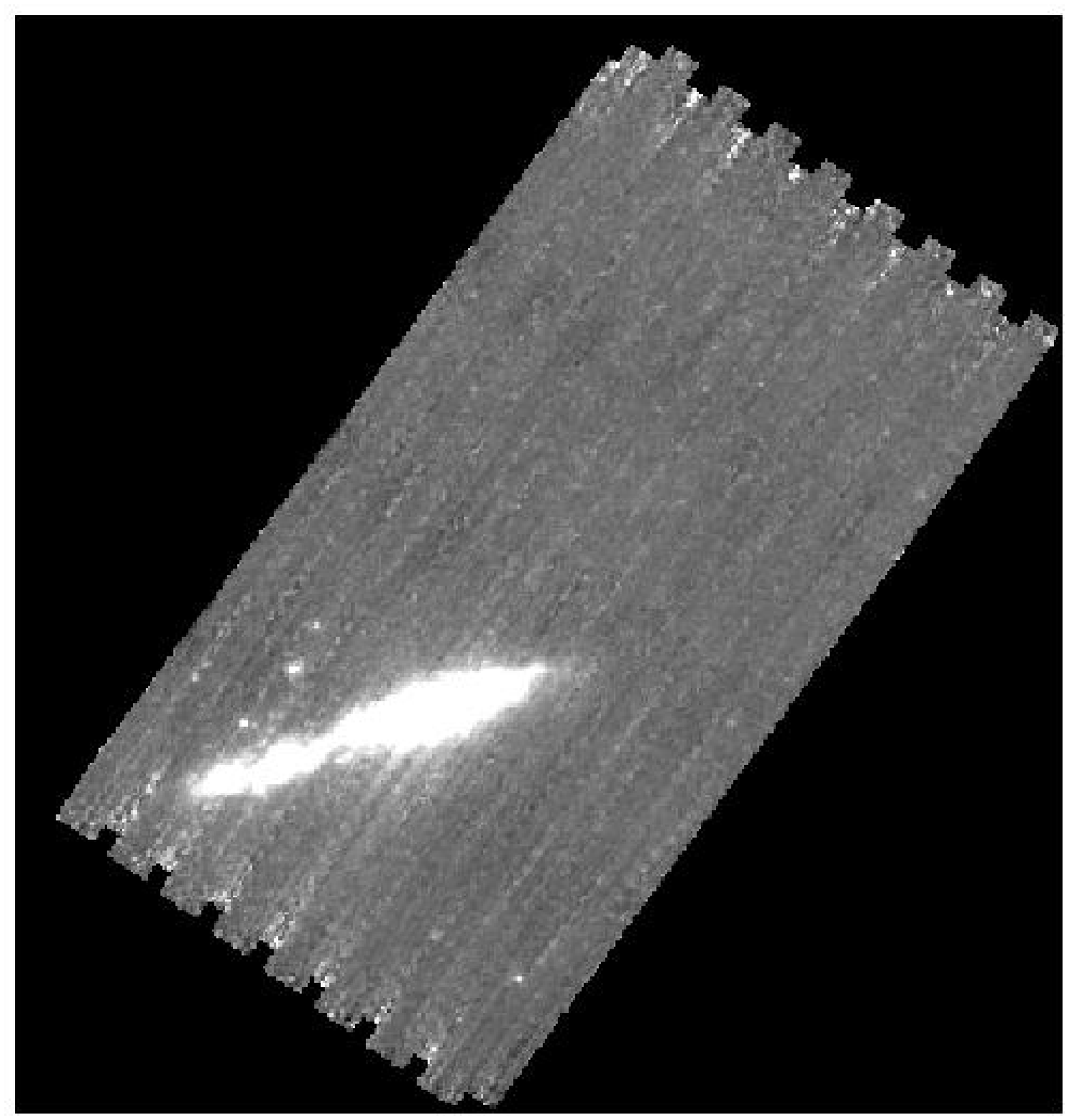}
\end{tabular}
\end{center}
\caption[ngc55] 
{ \label{fig:ngc55_ex} The scan map mode observation of NGC~55 is
shown to illustrate the fast/slow response and stim flash latent
temporal variations.  This observation consists of 16 scan legs
covering an area of 30 by 60 arcminutes and are discussed at length in
Engelbracht et al.\cite{Engelbracht04}.  The left panel gives the scan
map without the fast/slow and stim flash latent corrections.  The
middle panel shows the map with the fast/slow response variation
correction applied.  The right panel shows the final image with the
stim flash latent and fast/slow response variation corrections
applied. }
\end{figure}

The operating parameters for the 70~$\mu$m array have been changed a
number of times during the first 6 months of the mission.  Half of the
70~$\mu$m array was lost due to a telescope cable failure when the
cryostat vacuum shell cooled below 28K. In addition, the working half
of the array was noisier than expected, mostly due to the increased
hit rate from cosmic rays and in particular cosmic rays causing very
large charge deposition (on-orbit hit rates are 2 - 3 times
expectations). The bias voltage of the array was initially increased
to attempt to overcome the higher noise levels.  After extensive
analysis of the first few months of data, it was found that a lower
array bias was better overall.  Although the nominal signal-to-noise
was found to be largely independent of detector bias, there were a
number of undesirable effects that increased rapidly when the detector
bias was raised above 45mV. They included a rapid increase in stim
latent amplitudes, and a slowly changing effect on the shape of the
stim latent signals.  The ratio of fast to slow response also appeared
to be unstable. By reducing the bias voltages to 40mV or below, these
effects were reduced to the levels seen in prelaunch laboratory
experiments. They are illustrated in Figure~\ref{fig:ngc55_ex}, data
obtained with the bias set to 62mV. We discuss these effects in more
detail below.

From preflight testing, it was known that the fast/slow response was
time dependent at something like the 5\% level.  That is, the
variation of the fast response as measured by the stim flashes was
different from the variation of the slow response by about 5\%.  No
correction step was formally envisioned previously, but the effect was
large enough at the high array bias level to require correction.  This
correction is possible during the ``Redundancy'' stage of reduction,
but for this to be possible two sets of observations are required, one
which is time reversed from the other.  As it has only been possible
to do this for scan maps recently and not possible to do this for
photometry data, the fast/slow response correction has only been
possible by hand on a case-by-case basis.

The preflight expectations for stim flash latents were that they would
be dependent on stim flash amplitude and background level.  For
routine operations, the stim flash amplitude is fixed.  At the high
bias setting, there was a quick time evolution of the stim flash
latents from basically non-existent right after an anneal to very
strong 3 hours after an anneal.  This behavior is much lower, but
still present, at the current lower array bias level.  Given possible
dependencies both on time since anneal and background level, an
automatic correction may not be possible.  If it is, it will require
significant development and analysis of data.  In the meantime, there
are two possibilities for dealing with the effect. In one, a high pass
filtering algorithm is used to artificially flatten the baseline,
including any residuals from stim flashes. However, this approach
cannot be used for extended source images, which are affected by the
filter.  As a result, correcting for stim flash latents on such
sources is a ``Hand Processing'' step that is dependent on the mode
(photometry versus scan map) and target (degree of source extent) of
the data.

The nonlinearities in the Ge arrays are split into those caused by the
electronics and those caused the detector material which are dependent
on the flux level.  The electronic nonlinearities were measured in
preflight tests, but similar flight tests have shown some differences.
As such, the electronic nonlinearity correction awaits a new
calibration before being implemented.  The flux nonlinearity
calibration is composed of two parts.  The first uses the stimulator
flashed at different levels on different backgrounds to correct for
the pixel-to-pixel variations in the flux nonlinearities.  The second
uses measurements of calibration stars to remove the global flux
nonlinearities.  The total characterization of the flux nonlinearities
is a complex process and is ongoing.  As such, it has not been
implemented yet.

\begin{figure}
\begin{center}
\begin{tabular}{c}
\includegraphics[height=6cm]{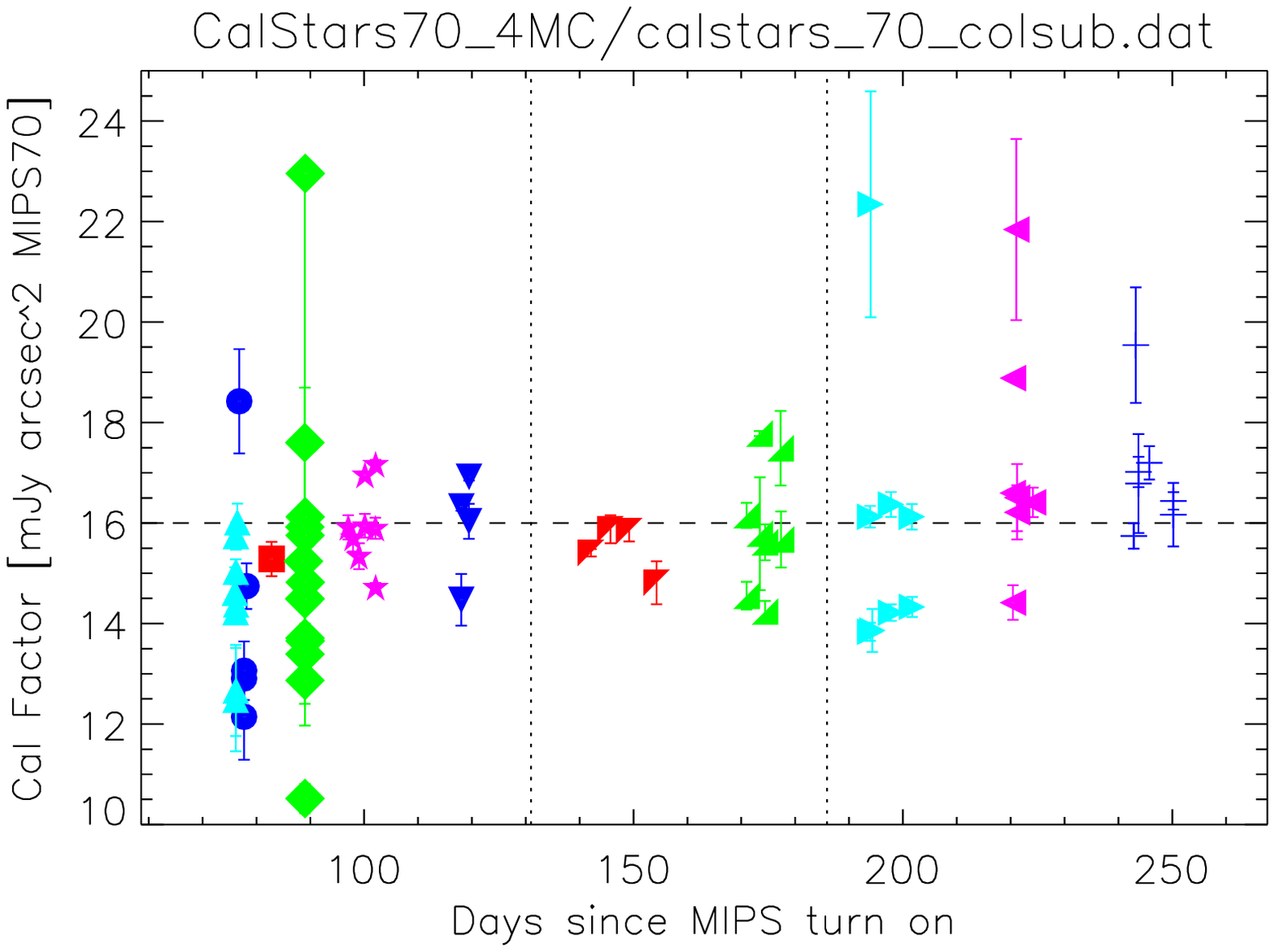}
\includegraphics[height=6cm]{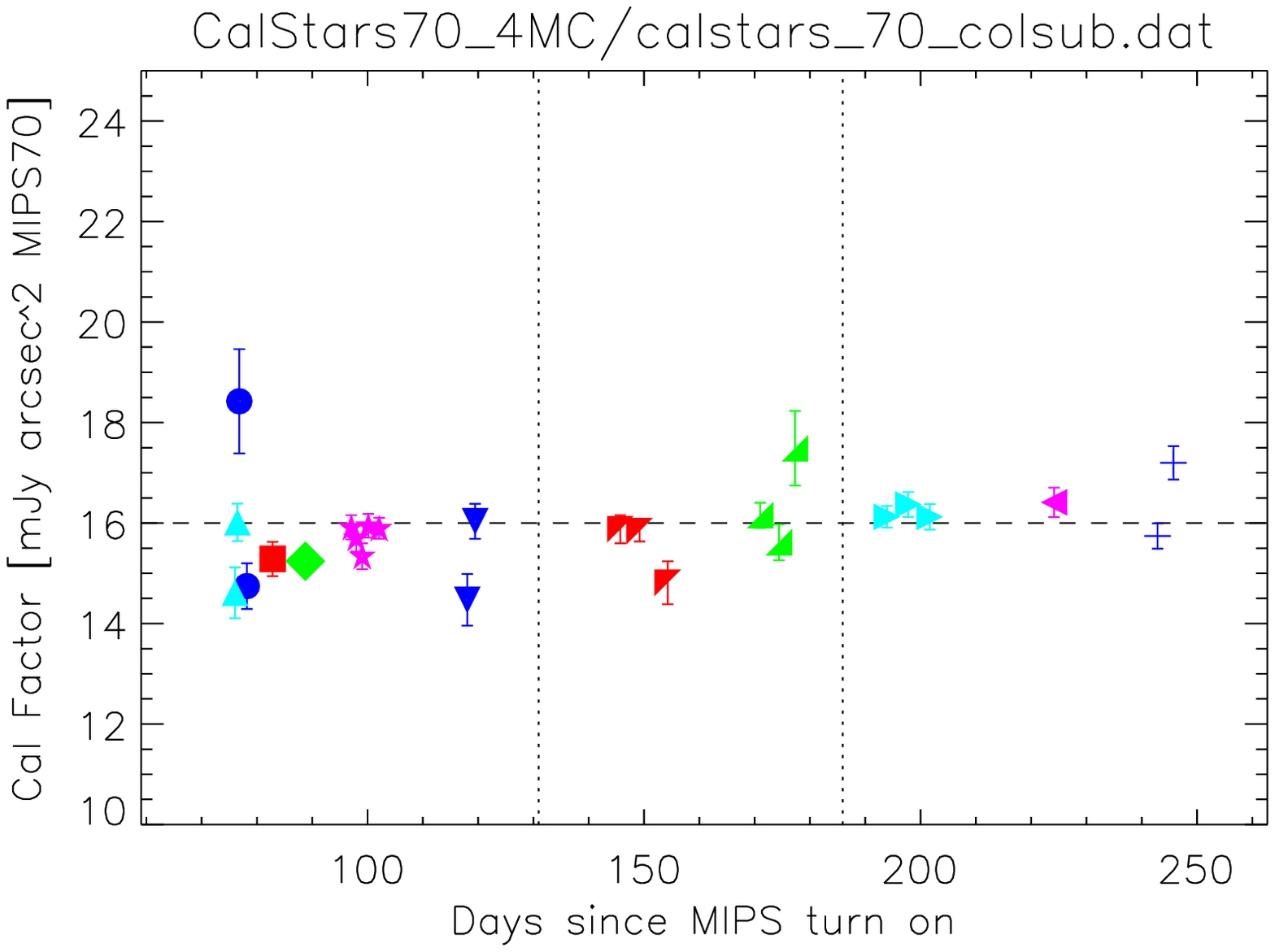}
\end{tabular}
\end{center}
\caption[calfactor_70] 
{ \label{fig:calfactor_70} The 70~$\mu$m calibration factor measured
for all the calibration star measurements (left) and for just
HD~163588 (right).  The error bars on each point are determined from
the sky noise in each individual measurement.  The first vertical line
denotes where the photometry mode dither pattern was changed and the
second vertical line denotes where the array bias was changed from 62
to 35~mV.  The horizontal line is drawn at 16 mJy arcsec$^{-2}$
(MIPS70 unit)$^{-1}$ (not the average) as a guide to the eye.}
\end{figure} 

The 70~$\mu$m absolute calibration is based on solar analog, A, and K
giant stars.  Using all the calibration star measurements taken to
date, the accuracy of 70~$\mu$m measurements is around 10\% (see
Figure~\ref{fig:calfactor_70}).  These calibration measurements
include stars with predicted fluxes ranging from 30~mJy to 5~Jy on
backgrounds between 4 and 25~MJy/sr.  One of the calibration stars
(HD~163588) is measured every MIPS campaign and the repeatability of
this star is better than 4\%.  The main difference between these two
accuracies is likely the result of the nonlinearities still to be
corrected.

\subsection{160~$\mu$m}

The 160~$\mu$m array is working well and the preflight reduction
algorithms are effective.  The largest change for 160~$\mu$m has been
the discovery in flight that this array suffers from a blue ``leak''
(actually an unexpected scattered reflection).  For a stellar source, this
``leak'' is about 15 times larger than the 160~$\mu$m stellar flux
making it difficult to use stars for calibration.  For sources with
redder than stellar flux distributions, the leak is usually below the
confusion noise. 

Unlike the 70~$\mu$m array, the 160~$\mu$m stim flash latents are
basically nonexistent after the first few seconds following the stim
flash.  This is different from ground test data where the stim flashes
lasted long enough to warrant correction.   This may be due to the
lower bias in flight than in ground testing and/or the high cosmic ray
rate (about 1 cosmic ray every 11 or so seconds).  Thus, the stim
flash latent correction step has been removed entirely for 160~$\mu$m
data.

As with the 70~$\mu$m array, the nonlinearities at 160~$\mu$m are not
currently characterized well enough to allow correction.  The
electronic nonlinearities are different than seen in ground tests.
The flux nonlinearities await sufficient asteroid data to allow for
the global flux nonlinearities to be calibrated.

Unlike the other two MIPS arrays, the calibration of the 160~$\mu$m
data is based on a variety of measurements that include asteroids,
cold sources that are well-measured by other space missions (ISO,
COBE), and objects where an accurate flux level can be determined by
interpolation and modeling such as stellar debris disks. The resulting
multiple calibrations agree to within 20\%. However, given that these
were not the planned calibration sources, the 160~$\mu$m flux
calibration is currently more uncertain than for the other two MIPS
arrays. Improvements in the calibration will depend largely on
measurements of asteroids.  The modeling of asteroids is more involved
than that of stars. To constrain models, each asteroid is observed at
both 70 and 160~$\mu$m and the 70~$\mu$m measurement is entered as a
constraint in the modeling.  The accuracy of the 160~$\mu$m
calibration is expected to improve significantly as more asteroids are
observed and the modeling is improved.

%%%%%%%%%%%%%%%%%%%%%%%%%%%%%%%%%%%%%%%%%%%%%%%%%%%%%%%%%%%%%
\acknowledgments     %>>>> equivalent to \section*{ACKNOWLEDGMENTS}       
This work was supported by NASA JPL contract 960785.

%%%%%%%%%%%%%%%%%%%%%%%%%%%%%%%%%%%%%%%%%%%%%%%%%%%%%%%%%%%%%
%%%%% References %%%%%

\bibliography{spie_2004_refs} %>>>> bibliography data in report.bib
\bibliographystyle{spiebib}   %>>>> makes bibtex use spiebib.bst

\end{document}